\documentclass[aps,prl,twocolumn,showpacs,superscriptaddress]{revtex4}  
\usepackage{graphicx}  
\usepackage{bm}        
\usepackage{amssymb}   
\usepackage{amsmath}
\usepackage[latin1]{inputenc}

\newcommand{\ket}[1]{\left| #1\right\rangle}

\newcommand{\mean}[1]{\langle #1\rangle}

\begin{document}
\title{Theory of the strong coupling between quantum emitters and propagating surface plasmons}
\author{A. Gonz\'alez-Tudela}
\affiliation{Departamento de F\'{\i}sica Te\'orica de la Materia Condensada and Condensed Matter Physics Center (IFIMAC), Universidad Aut\'onoma de Madrid, 28049, Spain.}
\author{P. A. Huidobro}
\affiliation{Departamento de F\'{\i}sica Te\'orica de la Materia Condensada and Condensed Matter Physics Center (IFIMAC), Universidad Aut\'onoma de Madrid, 28049, Spain.}
\author{L. Mart\'in-Moreno}
\affiliation{Instituto de Ciencia de Materiales de Arag\'on and Departamento de F\'isica de la Materia Condensada, CSIC-Universidad de Zaragoza, E-50009, Zaragoza, Spain}
\author{C. Tejedor}
\email[Corresponding author: ]{carlos.tejedor@uam.es}
\affiliation{Departamento de F\'{\i}sica Te\'orica de la Materia Condensada and Condensed Matter Physics Center (IFIMAC), Universidad Aut\'onoma de Madrid, 28049, Spain.}
\author{F.J. Garc\'ia-Vidal}
\email[Corresponding author: ]{fj.garcia@uam.es}
\affiliation{Departamento de F\'{\i}sica Te\'orica de la Materia Condensada and Condensed Matter Physics Center (IFIMAC), Universidad Aut\'onoma de Madrid, 28049, Spain.}
\date{\today}

\begin{abstract}
Here we present the theoretical foundation of the strong coupling phenomenon between quantum emitters and propagating surface plasmons observed in two-dimensional metal surfaces. For that purpose, we develop a quantum framework that accounts for the coherent coupling between emitters and surface plasmons and incorporates the presence of dissipation and dephasing. Our formalism is able to reveal the key physical mechanisms that explain the reported phenomenology and also to determine the physical parameters that optimize the strong coupling. A discussion regarding the classical or quantum nature of this phenomenon is also presented.
\end{abstract}

\maketitle

Surface plasmon polaritons (SPPs), hybrid bound modes comprising both electromagnetic fields and charge currents, are well-known to have both a subwavelength confinement and propagation lengths of tens or even hundreds of wavelengths \cite{barnes03,Maier07}. For this reason, the interaction between quantum emitters (QEs) and SPPs has attracted great interest recently \cite{Novotny06,trugler08a,Lodahl}. It has been shown that QE-SPP coupling can lead to single SPP generation \cite{akimov07a,chang07a,huck11a} and that the interaction between two QEs can be mediated by SPPs, resulting in energy transfer, superradiance \cite{martincano10a} and entanglement phenomena \cite{Dzsotjan,gonzaleztudela11a,martincano11}. Recently,  there have also been several experimental studies that show the emergence of strong coupling (SC), i.e., coherent energy exchange between propagating SPPs and  excitons either in organic molecules \cite{bellessa04a,dintinger05a,Salomon09,Schwartz,Aberra12,hakala09a} or in quantum dots \cite{vasa08a,bellesa08,Gomez10}. However, up to our knowledge, a first-principles explanation of these experimental results has not been presented yet.  

In this Letter we analyze the phenomenon of SC between quantum emitters (or absorbers) and SPPs and present its theoretical foundation. We develop a complete quantum treatment that is able not only to calculate absorption spectra and reproduce the experimental phenomenology, but also to deal with more complex aspects as photon statistics.  

\begin{figure}[htbp]
\centering
\includegraphics[width=\columnwidth]{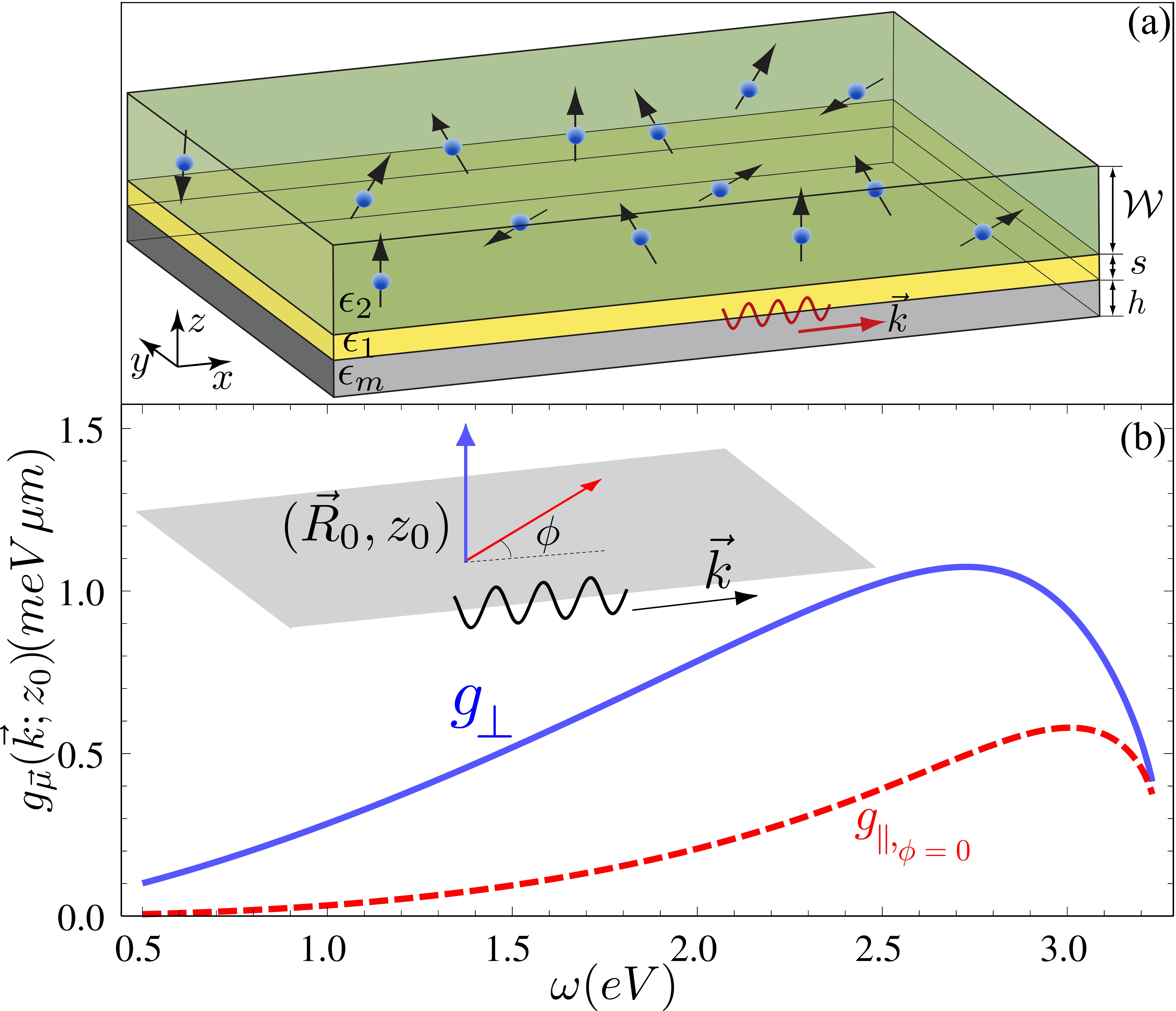}
\caption{(color online) (a) Schematic picture of the $N$ QEs distributed in a volume of width $\mathcal{W}$ separated by a distance $s$ from a metal film of thickness $h$. (b) Coupling constant, $g_{\vec{\mu}}(\vec{k};z_0)$, for a single QE with perpendicular (solid blue) and parallel (red dashed) orientations (see inset), placed at $z_0=20$nm and interacting with a SPP of momentum $\vec{k}(\omega)$.}
\label{fig1}
\end{figure}

In Figure 1(a) we render a sketch of the general structure that mimics the experimental configuration: a collection of $N$ QEs immersed into a layer of thickness $\mathcal{W}$, and placed on top of a thin metal film (thickness $h$). In this work the acronym QE will refer to a quantum system with discrete electronic levels, like organic molecules or quantum dots. In some of the experimental set-ups and in order to avoid quenching of the QEs, a dielectric spacer of width $s$ is located between the QEs and the metal substrate. We will take $\epsilon_2=\epsilon_1=1$ in our calculations and we will use the dielectric function of the metal (silver), $\epsilon_m$, as tabulated in Ref. \cite{Palik}. As a minor simplification, we will assume a semi-infinite metal substrate instead of the metal film considered in the experiments (these films are thick enough for the SPPs to be very similar to those of a single interface). Each QE is represented by a two-level system (2LS) $\{\ket{g},\ket{e}\}$ and characterized by a transition frequency $\omega_0$ (in this paper we will use $\omega_0=2\,$eV, $\hbar=1$) and dipole moment $\vec{\mu}$ with spontaneous decay rate $\gamma_0=\omega_{0}^{3}\mu^2/(3\pi\epsilon_0 c^3)$. This description assumes a large separation between the electronic energy levels of the emitter, with only one possible transition at the excitation frequency.  

A QE placed in the vicinity of a metal surface can decay into three different channels \cite{barnes98a}: excitation of SPPs that propagate along the metal surface, radiation of photons into the far field and dissipation through ohmic losses in the metal. In order to study the SC regime between the QE and SPPs, the excitation of plasmons will be considered as the coherent channel, while radiation to free space and losses into the metal will be treated as dissipation mechanisms. The general Hamiltonian that describes the coherent interaction between $N$ QEs and the 2D-SPPs can be written as (a detailed account of its derivation is presented in the Supplementary Material):
\begin{eqnarray}
\label{Eqsinglemitterconti}
H^{N}&=&\sum_{j=1}^{N_L}\sum_{i=1}^{N_s} \omega_0 \sigma^{\dagger}_{i,j}\sigma_{i,j}+\sum_{\vec{k}}\omega(\vec{k}) a^{\dagger}_{\vec{k}} a_{\vec{k}} + \nonumber \\ & & 
\sum_{\vec{k}}\sum_{j=1}^{N_L}\sum_{i=1}^{N_s} [\frac{g_{\vec{\mu}}(\vec{k};z_j)}{\sqrt{A}}a_{\vec{k}}\sigma_{i,j}^{\dagger}e^{i\vec{k}\vec{R}_i}+c.c.],
\end{eqnarray}
where $\sigma_{i,j}$ and $\sigma^{\dagger}_{i,j}$ are the QE lowering and raising operators of a QE that is located at $(\vec{R}_i,z_j)$. In our modeling we assume the ensemble of QEs to be disposed in $N_L$ layers, each of them having $N_s$ equal emitters such that $N=N_L \times N_s$.  In Eq.(1), $a_{\vec{k}}$ and $a^{\dagger}_{\vec{k}}$ are the destruction and creation operators for the SPP quantum field with in-plane momentum $\vec{k}$ and energy $\omega$, linked by the dispersion relation $k^2(\omega)(\epsilon_m+1)=\epsilon_m \omega^2/c^2$. The area of the metal-dielectric interface is $A$ and $g_{\vec{\mu}}(\vec{k};z_j)$ is the coupling constant of the dipolar interaction between a given QE and the SPP field:
\begin{equation}
g_{\vec{\mu}}(\vec{k};z)=\sqrt{\frac{\omega(\vec{k})}{2 \epsilon_0 L(\vec{k})}} e^{-k_z z} \vec{\mu} \cdot  (\hat{u}_{\vec{k}}+i\frac{|\vec{k}|}{k_z}\hat{u}_z),
\end{equation}
\noindent where $L(\vec{k})$ is the {\it effective} length of the mode \cite{Zayats,Greffet}. For calculating this coupling constant, propagation losses of the SPP modes are neglected. The unitary vectors in the $\vec{k}$ and $z$ directions are $\hat{u}_{\vec{k}}$ and $\hat{u}_z$, respectively. The dependence of $g_{\vec{\mu}}$ with $z$ is dictated by the decay length of the SPP in the z-direction via $k_z=\sqrt{k^2-\omega^2/c^2}$. In Fig. 1(b), we render the evolution of $g_{\vec{\mu}}(\vec{k};z)$ with frequency for two possible orientations of the dipole: parallel to the momentum $\vec{k}$ and perpendicular to the metal surface. In both cases, the couplings are evaluated for QEs with  $\gamma_0=0.1$meV, which is a typical value for the J-aggregates used in the experiments as QEs \cite{bellessa04a,dintinger05a,Salomon09,Aberra12}. As shown in Fig.1(b), the coupling constant between the QE and the SPP mode is larger for the perpendicular orientation, as $k_z$ is always smaller than $|\vec{k}|$.    

To simplify the general Hamiltonian (1), we first consider that, in the low excitation regime, the QE lowering and raising operators  ($\sigma_{i,j}$ and $\sigma^{\dagger}_{i,j}$), can be replaced by bosonic operators, $b_{i,j}$ and $b^{\dagger}_{i,j}$, respectively. Second, as in the experiments the ensemble of QEs is disordered, we assume that the structure factor is peaked at zero-momentum. Third, we build up a collective mode of the $N$ QEs, $D^{\dagger}_{\vec{k}}$, by means of a transformation in which each excitation is weighted by its coupling to SPPs.  Based on this, the total hamiltonian of the $N$ QEs interacting with the SPP modes of a 2D-metal film can be written as $H^N=\sum_{\vec{k}}H^{N}_{\vec{k}}$ (see the details of its derivation in the Supplementary Material), in which the hamiltonian associated with momentum $\vec{k}$ has the following expression:
\begin{equation}
H^{N}_{\vec{k}}=\omega_0 D^{\dagger}_{\vec{k}}D_{\vec{k}}+\omega(\vec{k}) a^{\dagger}_{\vec{k}} a_{\vec{k}}+[g^N_{\vec{\mu}}(\vec{k})a_{\vec{k}}D^{\dagger}_{\vec{k}}+c.c.].
\end{equation}

Here, $g^N_{\vec{\mu}}(\vec{k})$ is the effective coupling constant:
\begin{equation}
g^N_{\vec{\mu}}(\vec{k})=\sqrt{\frac{N_s}{A}\sum_{j=1}^{N_L}|g_{\vec{\mu}}(\vec{k};z_j)|^2}=\sqrt{n \int_{s}^{s+\mathcal{W}}|g_{\vec{\mu}}(\vec{k};z)|^2 dz}.
\end{equation}  

The last equality in Eq. (4) assumes a continuum of layers in the $z$-direction with a total thickness $\mathcal{W}$ and a volume density of emitters $n=N_sN_L/(A\mathcal{W})$. The hamiltonian as written in Eq. (3) is one of the main results of our work, as it allows an {\it ab-initio} quantum treatment of the coherent coupling between an ensemble of $N$ QEs and SPPs. Notice that this interaction conserves the total momentum of the system composed of the supermode of QEs and the SPP. When evaluating the coupling constant for a momentum $\vec{k}$, $g^{N}_{\vec{\mu}}(\vec{k})$, there is no need to rely on fitting parameters and can be calculated from first principles, as shown below. 

\begin{figure}[htbp]
\centering
\includegraphics[width=\linewidth]{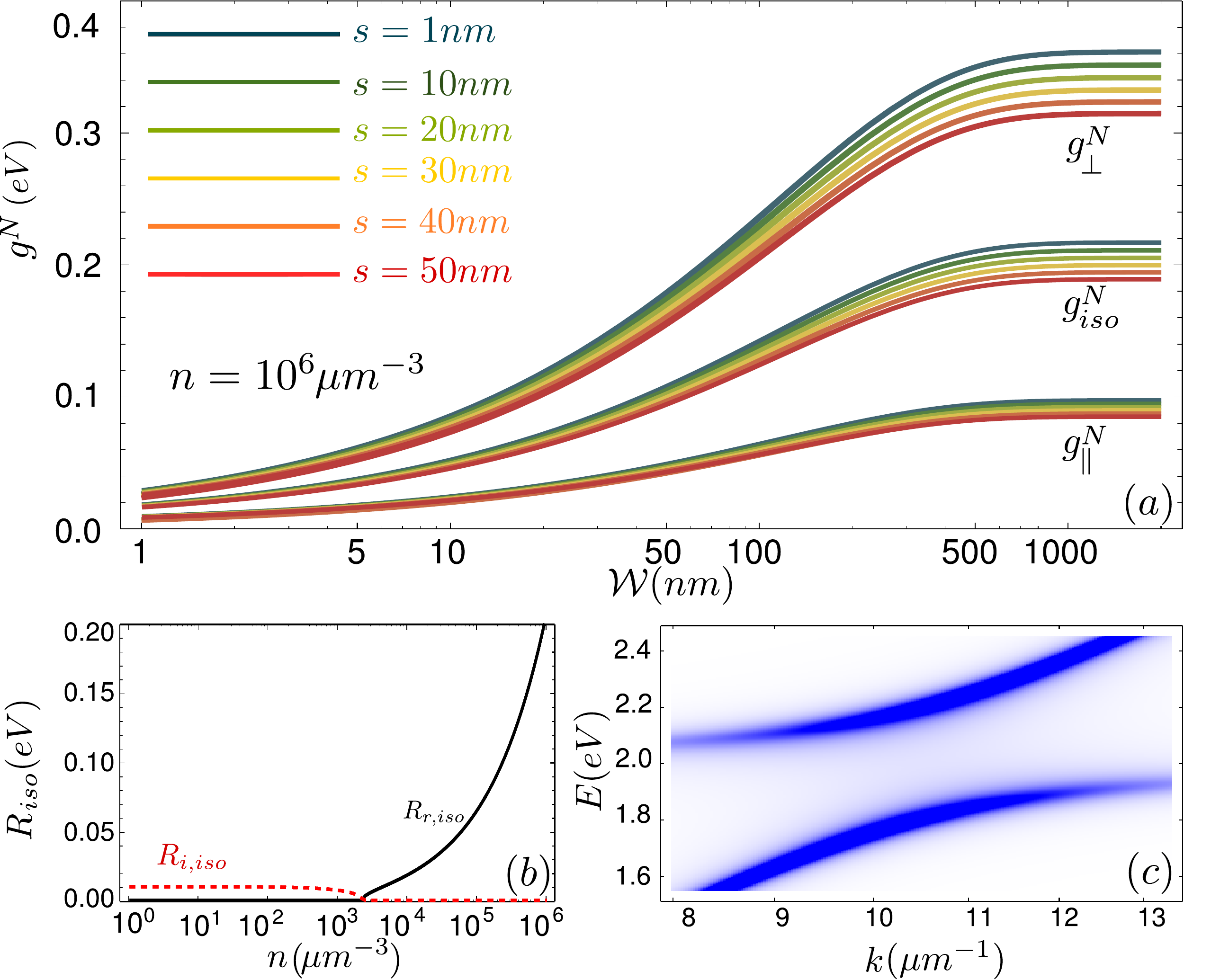}
\caption{(color online). (a) Coupling constant, $g^N(\mathcal{W})$, for separations $s$ ranging from $1$ to $50\,$nm and for parallel, perpendicular and isotropic orientations of the QEs with $\gamma_0=0.1\,$meV. (b) Real (solid black) and imaginary (red dashed) parts of the Rabi splitting at resonance for dipoles oriented isotropically, $R_{iso}$, as a function of $n$ for the geometrical parameters: $s=1$nm and $\mathcal{W}=500\,$nm and $\gamma_\phi=40\,$meV. (c) Polariton population (see main text) of a distribution of QEs as a function of $\vec{k}$, with the same geometrical parameters as in panel (b) and with $\Omega_{\vec{k}}=0.1 g^N$. The volume density in this case is $n=10^6 {\mu m}^{-3}$, as in panel (a).}
\label{fig3}
\end{figure}

In Figure 2(a) we plot the effective coupling constant $g^N$ evaluated at $\vec{k}(\omega_0)$ for a density of emitters $n=10^6 \mu$m$^{-3}$ (of the order of the volume densities used in the experiments) as a function of $\mathcal{W}$ and for different values of the spacer width. This magnitude depends on the orientation of the QEs' dipole moments. Here we render the two limiting cases (all dipoles oriented perpendicularly or parallel to the metal surface) as well as an isotropic average over these two orientations, $g_{iso}^2=2g_{\parallel}^2/3+g_{\perp}^2/3$. Two main conclusions can be extracted from this figure. First, $g^N$ depends strongly on $\mathcal{W}$ but saturates for thick enough films. This saturation is due to the exponential dependence of $g_{\vec{\mu}}$ on $z$, related to the spatial decay of the SPP mode, and, therefore, is determined by the dielectric environment of the metal film. Second, the dependence of $g^N_{\vec{\mu}}(\vec{k})$ on the width of the spacer layer is not very strong. 

The excitation of the hybrid system needs to be included into the theoretical framework. In order to reproduce the typical experimental configuration, we will assume that SPPs are excited by a coherent laser field. A new term is incorporated into the total Hamiltonian, $H^L_{\vec{k}}(t)=\Omega_{\vec{k}}(a_{\vec{k}}e^{i\omega_L t}+a^{\dagger}_{\vec{k}}e^{-i\omega_L t})$ \cite{gonzaleztudela11a}, in which $\Omega_{\vec{k}}$ measures the intensity of the laser field and $\omega_L$ is the operating frequency of the laser. In this way, the laser field fixes the SPP parallel momentum, $\vec{k}$, implying that only the term $H^{N}_{\vec{k}}$ in the total hamiltonian $H^{N}$ needs to be taken into account.

Finally, the description of the dynamics of the system must be completed by considering both the losses in the ensemble of QEs and the dissipation associated with the SPP mode.  The decay lifetime of the SPP mode, $\gamma_{a_{\vec{k}}}$, can be calculated from the SPP propagation length, $L_{\mathrm{SPP}}$, and group velocity, $v_{\mathrm{g}}$, $\gamma_{a_{\vec{k}}}=v_{\mathrm{g}}/ L_{\mathrm{SPP}}$. This SPP lifetime increases as the frequency approaches the SPP cut-off frequency, being around $5\,$meV for $\omega=\omega_0=2\,$eV. The lifetime associated with the collective mode of the ensemble of $N$ QEs, $\gamma_{D_{\vec{k}}}$, is obtained from the averaged value of the decay rates for each individual QE, $\gamma_{\sigma}(z)$, weighted by a term proportional to $|g_{\vec{\mu}}(\vec{k},z)|^2$ (for details see Supplementary Material). Additionally, in order to be as close as possible to the experimental conditions, the existence of vibro-rotational states in organic molecules must also be taken into account. These degrees of freedom within the QEs can be incorporated into the 2LS model by means of pure dephasing mechanisms, characterized by a dephasing rate, $\gamma_{\phi}$. In this work we take $\gamma_\phi=40\,$meV \cite{Fidder90}, which is a typical value at room temperature for the organic molecules used to observe SC between $N$ QEs and SPPs. 

With all these ingredients, we use a Markovian master equation for the density matrix and introduce perturbatively the corresponding Lindblad operators \cite{carmichael02} associated with each of the three dissipative channels. Recalling that the general expression of a Lindblad term associated with an arbitrary operator $c$ is $\mathcal{L}_c=(2 c\rho c^\dagger- c^\dagger c\rho-\rho c^\dagger c)$, the master equation for the density matrix associated with momentum $\vec{k}$, $\rho_{\vec{k}}(t)$, can be written as:
\begin{equation}
\dot{\rho}_{\vec{k}}=i [\rho_{\vec{k}},H^{N}_{\vec{k}}+H^L_{\vec{k}}]+\frac{\gamma_{D_{\vec{k}}}}{2}\mathcal{L}_{D_{\vec{k}}}+\frac{\gamma_{a_{\vec{k}}}}{2}\mathcal{L}_{a_{\vec{k}}}+\frac{\gamma_{\phi}}{2}\mathcal{L}_{D^{\dagger}_{\vec{k}}D_{\vec{k}}}.
\end{equation} 

The solution of the master equation for $\vec{k}_0=\vec{k}(\omega_0)$ (in-plane momentum that displays maximum coupling) yields to coherence functions being proportional to $\exp(iRt)$ where $R$ is the Rabi splitting at resonance:
\begin{equation}
R=\sqrt{[g^{N}_{\vec{\mu}}(\vec{k}_0)]^2-({\gamma_{D_{\vec{k}_0}}+\gamma_{\phi}-\gamma_{a_{\vec{k}_0}}})^2/16}.
\end{equation}

Following the standard analysis \cite{laussy08a}, we will consider that our hybrid system is within the SC regime when the imaginary part of the Rabi splitting is zero. In Fig. 2(b), we plot the evolution of $R \equiv R_r+iR_i$ with the volume density $n$ for an ensemble of N QEs whose dipoles are oriented isotropically.  For very low densities (for this set of parameters, $n<2 \times 10^3 {\mu m}^{-3}$), $R$ is a purely imaginary number and, therefore, the system operates in the WC regime. This density threshold, $n_t$, is mainly controlled by $\gamma_\phi$ as $\gamma_\phi \gg \gamma_D, \gamma_a$ for this set of decay rates. Notice that as $\gamma_\phi$ decreases exponentially when lowering the temperature \cite{Fidder90},  $n_t$ is expected to be much smaller at very low temperatures (by assuming $\gamma_\phi=0$ at zero temperature, $n_t$ would be around $20\, {\mu m}^{-3}$). For high enough densities  ($n \approx 10^5-10^8 {\mu m}^{-3}$, typical densities in the experiments \cite{hakala09a}), $R_r$ (the so-called vacuum Rabi splitting) is dominated by the coupling constant $g^{N}$ as $g^{N} \gg \{\gamma_D,\gamma_a,\gamma_\phi\}$ and $R_r \approx g^{N}$. As this coupling constant scales as $\sqrt{n}$, so does $R_r$, as observed in the experiments. Within our formalism, it is also possible to evaluate the absorption spectra, a magnitude that is attainable experimentally. In Fig. 2(c) we plot the polariton population (the sum of both the QEs supermode and SPP mode occupations, a magnitude that is proportional to the absorption by the system \cite{loudon_book00a}) versus energy and parallel momentum, showing the anti-crossing between the flat band at $\omega_0$ associated with the collective mode of the $N$ QEs and the dispersive band of the SPPs.  Already existent experimental results  \cite{hakala09a} can be confronted with our theoretical framework. In that experiment, the metal film was silver, $\mathcal{W} = 50$nm and an ensemble of $n =1.2 \times 10^8 \mu$m$^{-3}$ Rodhamine 6G molecules were used as QEs ($\gamma_0=1\mu$eV, as reported in Ref. \cite{ambrose94a}). This resulted in the observation of a Rabi splitting of $0.115$ eV. For those parameters, our theory predicts $R_r=0.04$ eV for parallel-oriented QEs, $R_r= 0.18$ eV for the perpendicular orientation and $R_r=0.10$ eV for an isotropic average, showing a good agreement between theory and experiment.

Now we address the fundamental question regarding the classical/quantum nature of the SC regime observed in this type of systems. Although a semiclassical formalism fed with phenomenological parameters is able to reproduce qualitatively the reported absorption spectra \cite{Salomon12}, this should not be taken as a statement that the system contains no interesting quantum physics. Non-classicality is unambiguosly revealed by the presence of photon antibunching in the dynamics of the strongly coupled system. For this reason, we analyze the behavior of the second-order correlation function, $g^{(2)}$, defined as $g^{(2)}(\tau)=\lim_{t\rightarrow \infty}\mean{{D}^{\dagger}_{\vec{k}}(t)({D}^{\dagger}_{\vec{k}}D_{\vec{k}})(t+\tau)D_{\vec{k}}(t)}/\mean{{D}^{\dagger}_{\vec{k}}D_{\vec{k}}(t)}^2$. Photon antibunching yields $g^{(2)}(0)<1$. However, within the approximations leading to Hamiltonian (3) with $D_{\vec{k}}$ constructed from bosonic operators, the system behaves as two coupled harmonic oscillators. In this case, $g^{(2)}(0)$ is always greater or equal to $1$ \cite{mandel_book95a} and its time evolution critically depends on the excitation means. For the case of coherent pumping, the system acquires the statistics of the laser field and hence $g^{(2)}(\tau)=1$. The case of incoherent pumping is simulated in our theoretical formalism by introducing a Lindblad term, $P_{D_k}\mathcal{L}_{{D}^{\dagger}_{\vec{k}}}/2$ \cite{laussy08a}, into the master equation [Eq.(5)], instead of $H^L_{\vec{k}}$. As shown in the inset of Fig. 3(b) (green-dashed line), when the collective mode is driven incoherently $g^{(2)}(0)=2$ and its time dependence presents some Rabi oscillations but $g^{(2)}(\tau) \ge 1$.

In order to find fingerprints of non-classicality in our system ($g^{(2)}(0) < 1$), it is necessary to incorporate a non-linear term into Hamiltonian (3):
\begin{equation}
H_{nl}=\sum_{\vec{k},{\vec{k}}^{\prime},\vec{q}}U_{D}{D}^{\dagger}_{\vec{k}+\vec{q}}{D}^{\dagger}_{\vec{k'}-\vec{q}}D_{\vec{k}}D_{\vec{k'}}\,.
\end{equation}

The physical origin of this term can be twofold  \cite{ciuti00a}: a direct coupling between the QEs, similar to the coulomb interaction between excitons reported in semiconductor structures, and/or saturation effects. In this last case we can even quantify this contribution by introducing the second-order correction within the Holstein-Primakoff approach \cite{holstein40a} in the process of replacing the QE lowering and raising operators by the bosonic ones. By considering a quasi-2D layer of QEs, a non-linear term as expressed by Eq. (7) can be straightforwardly obtained from the general Hamiltonian (1) with $U_D$ being $-\omega_0/N$ (technical details are given in the Supplementary Material). Notice that whereas in the {\it linear} case the key parameter is the volume density $n$, the saturation contribution to the non-linear term is controlled by the total number of active QEs, $N$. In Fig. 3(a), we show the dependence of $g^{(2)}(0)$ on $|U_D|$ and the frequency detuning, $(\omega_0-\omega_L)$, both expressed in units of $g^{N}$. In these calculations we have taken a dephasing rate $\gamma_\phi=0$ in order to find the most favorable, yet experimentally feasible, conditions to observe photon antibunching.  As we consider pumping to only one $\vec{k}$-state, the population of a SPP-mode with parallel momentum $\vec{k}^{\prime}$ is proportional to $\delta(\vec{k}-\vec{k}^{\prime})$, canceling out the summation in $\vec{k}^{\prime}$ in Eq.(7). In addition, the summation in $\vec{q}$ can also be neglected because the shape of the SPP dispersion relation does not allow parametric scattering  \cite{ciuti01a,verger06a}, in which both energy and momentum are conserved, to SPP states with $\vec{q} \neq 0$. Two particular cases ($|U_D|=0.025g^N$ and $0.005g^N$) are displayed in Fig. 3(b) for a better visualization. If we assumed a saturation origin for $U_D$, these two cases would correspond to $N \approx 2\times 10^3$ and $N \approx 10^4$, respectively. Importantly, photon antibunching is observed in both cases and is greater when the laser frequency almost coincides with $\omega_0$ or $\omega_0 \pm g^N$. Therefore, our results suggest that in order to observe noticeable photon antibunching, the experiments should be performed at very low temperature to avoid dephasing and plasmon losses. Additionally, in order to reduce the number of active QEs, the laser beam should have a very small spot size, and the QEs should be disposed forming quasi-2D layers.

\begin{figure}[htbp]
\centering
\includegraphics[width=\linewidth]{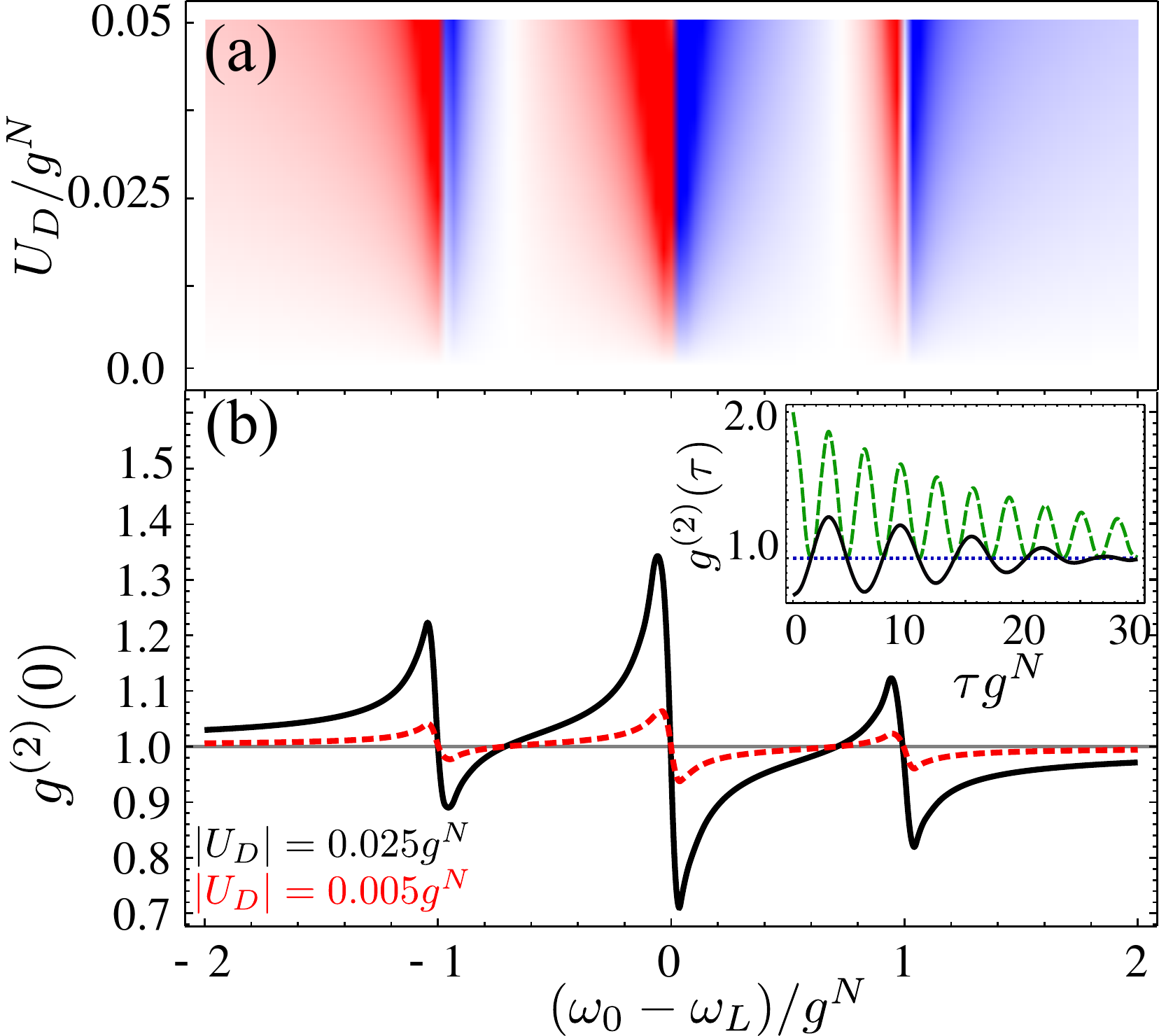}
\caption{(color online) (a) Contour plot of $g^{(2)}(0)$ as a function of the non-linearity, $|U_D|/g^N$, and detuning, $(\omega_0-\omega_{\mathrm{L}})/g^N$, for the coherently pumped configuration, with system parameters: $s=10 \,nm$, $\mathcal{W}=10 \,nm$ and $n=10^6\mu m^{-3}$, which yields $g^N\approx 50 \,meV$, see Fig. 2(a). The color code is: $0$ blue; $1$ white; $2$ red. (b)  Horizontal cuts of (a) at two fixed non-linear parameters: $|U_D|=0.005g^N$ (dashed red) and $0.025g^N$ (solid black). Inset: $g^{(2)}(\tau)$ for a system with $|U_D|=0.025g^N$ and $\omega_{\mathrm{L}}=\omega_0+0.1g^N$ (solid black) and with $|U_D|=0$ for both incoherent pumping (dashed green) and coherent excitation (dotted blue). In all these calculations we have taken a dephasing rate $\gamma_\phi=0$.}
\label{fig4}
\end{figure}

In conclusion, we have presented an {\it ab-initio} quantum formalism to study the phenomenon of SC between QEs and propagating SPPs in two-dimensional metal surfaces. Based on this formalism, we are able to predict the critical density where SC emerges for a given geometry and distribution of QEs, and to determine the optimal geometrical parameters that maximize SC. Our results show that, for experiments carried out at room temperature, QE and SPP losses play a minor role in the emergence of SC. Both coherent coupling between the QEs and SPPs and pure dephasing mechanisms determine the strength of the phenomenon in this case. Additionally, the development of this general quantum framework allows studying the fundamental nature (classical versus quantum) of this phenomenon by analyzing the conditions in which photon antibunching could be observed. 

Work supported by the Spanish MINECO (MAT2011-22997, MAT2011-28581-C02 and CSD2007-046-NanoLight.es) and 
CAM (S-2009/ESP-1503). A.G.-T. and P.A.-H acknowledge FPU grants (AP2008-00101 and AP2008-00021, respectively) from the Spanish Ministry of Education. This work has been partially funded by the European Research Council (ERC-2011-AdG Proposal No. 290981).

\vspace*{-0.5cm}


\bibliography{plasmondef}

\begin{thebibliography}{36}
\expandafter\ifx\csname natexlab\endcsname\relax\def\natexlab#1{#1}\fi
\expandafter\ifx\csname bibnamefont\endcsname\relax
  \def\bibnamefont#1{#1}\fi
\expandafter\ifx\csname bibfnamefont\endcsname\relax
  \def\bibfnamefont#1{#1}\fi
\expandafter\ifx\csname citenamefont\endcsname\relax
  \def\citenamefont#1{#1}\fi
\expandafter\ifx\csname url\endcsname\relax
  \def\url#1{\texttt{#1}}\fi
\expandafter\ifx\csname urlprefix\endcsname\relax\def\urlprefix{URL }\fi
\providecommand{\bibinfo}[2]{#2}
\providecommand{\eprint}[2][]{\url{#2}}

\bibitem[{\citenamefont{Barnes et~al.}(2003)\citenamefont{Barnes, Dereux, and
  Ebbesen}}]{barnes03}
\bibinfo{author}{\bibfnamefont{W.~L.} \bibnamefont{Barnes}},
  \bibinfo{author}{\bibfnamefont{A.}~\bibnamefont{Dereux}}, \bibnamefont{and}
  \bibinfo{author}{\bibfnamefont{T.~W.} \bibnamefont{Ebbesen}},
  \bibinfo{journal}{Nature} \textbf{\bibinfo{volume}{424}},
  \bibinfo{pages}{824} (\bibinfo{year}{2003}).

\bibitem[{\citenamefont{Maier}(2007)}]{Maier07}
\bibinfo{author}{\bibfnamefont{S.~A.} \bibnamefont{Maier}},
  \emph{\bibinfo{title}{Plasmonics: fundamentals and applications}}
  (\bibinfo{publisher}{Springer}, \bibinfo{year}{2007}).

\bibitem[{\citenamefont{Anger et~al.}(2006)\citenamefont{Anger, Bharadwaj, and
  Novotny}}]{Novotny06}
\bibinfo{author}{\bibfnamefont{P.}~\bibnamefont{Anger}},
  \bibinfo{author}{\bibfnamefont{P.}~\bibnamefont{Bharadwaj}},
  \bibnamefont{and} \bibinfo{author}{\bibfnamefont{L.}~\bibnamefont{Novotny}},
  \bibinfo{journal}{Phys. Rev. Lett.} \textbf{\bibinfo{volume}{96}},
  \bibinfo{pages}{113002} (\bibinfo{year}{2006}).

\bibitem[{\citenamefont{Tr\"ugler and Hohenester}(2008)}]{trugler08a}
\bibinfo{author}{\bibfnamefont{A.}~\bibnamefont{Tr\"ugler}} \bibnamefont{and}
  \bibinfo{author}{\bibfnamefont{U.}~\bibnamefont{Hohenester}},
  \bibinfo{journal}{Phys. Rev. B} \textbf{\bibinfo{volume}{77}},
  \bibinfo{pages}{115403} (\bibinfo{year}{2008}).

\bibitem[{\citenamefont{Andersen et~al.}(2011)\citenamefont{Andersen, Stobe,
  Sorensen, and Lodahl}}]{Lodahl}
\bibinfo{author}{\bibfnamefont{M.~L.} \bibnamefont{Andersen}},
  \bibinfo{author}{\bibfnamefont{S.}~\bibnamefont{Stobe}},
  \bibinfo{author}{\bibfnamefont{A.~S.} \bibnamefont{Sorensen}},
  \bibnamefont{and} \bibinfo{author}{\bibfnamefont{P.}~\bibnamefont{Lodahl}},
  \bibinfo{journal}{Nature Physics} \textbf{\bibinfo{volume}{7}},
  \bibinfo{pages}{215218} (\bibinfo{year}{2011}).

\bibitem[{\citenamefont{Akimov et~al.}(2007)\citenamefont{Akimov, Mukherjee,
  Yu, Chang, Zibrov, Hemmer, Park, and Lukin}}]{akimov07a}
\bibinfo{author}{\bibfnamefont{A.~V.} \bibnamefont{Akimov}},
  \bibinfo{author}{\bibfnamefont{A.}~\bibnamefont{Mukherjee}},
  \bibinfo{author}{\bibfnamefont{C.~L.} \bibnamefont{Yu}},
  \bibinfo{author}{\bibfnamefont{D.~E.} \bibnamefont{Chang}},
  \bibinfo{author}{\bibfnamefont{A.~S.} \bibnamefont{Zibrov}},
  \bibinfo{author}{\bibfnamefont{P.~R.} \bibnamefont{Hemmer}},
  \bibinfo{author}{\bibfnamefont{H.}~\bibnamefont{Park}}, \bibnamefont{and}
  \bibinfo{author}{\bibfnamefont{M.~D.} \bibnamefont{Lukin}},
  \bibinfo{journal}{Nature} \textbf{\bibinfo{volume}{450}}
  (\bibinfo{year}{2007}).

\bibitem[{\citenamefont{Chang et~al.}(2007)\citenamefont{Chang, S\o{}rensen,
  Hemmer, and Lukin}}]{chang07a}
\bibinfo{author}{\bibfnamefont{D.~E.} \bibnamefont{Chang}},
  \bibinfo{author}{\bibfnamefont{A.~S.} \bibnamefont{S\o{}rensen}},
  \bibinfo{author}{\bibfnamefont{P.~R.} \bibnamefont{Hemmer}},
  \bibnamefont{and} \bibinfo{author}{\bibfnamefont{M.~D.} \bibnamefont{Lukin}},
  \bibinfo{journal}{Phys. Rev. B} \textbf{\bibinfo{volume}{76}},
  \bibinfo{pages}{035420} (\bibinfo{year}{2007}).

\bibitem[{\citenamefont{Huck et~al.}(2011)\citenamefont{Huck, Kumar, Shakoor,
  and Andersen}}]{huck11a}
\bibinfo{author}{\bibfnamefont{A.}~\bibnamefont{Huck}},
  \bibinfo{author}{\bibfnamefont{S.}~\bibnamefont{Kumar}},
  \bibinfo{author}{\bibfnamefont{A.}~\bibnamefont{Shakoor}}, \bibnamefont{and}
  \bibinfo{author}{\bibfnamefont{U.~L.} \bibnamefont{Andersen}},
  \bibinfo{journal}{Phys. Rev. Lett.} \textbf{\bibinfo{volume}{106}},
  \bibinfo{pages}{096801} (\bibinfo{year}{2011}).

\bibitem[{\citenamefont{Martin-Cano et~al.}(2010)\citenamefont{Martin-Cano,
  Martin-Moreno, Garcia-Vidal, and Moreno}}]{martincano10a}
\bibinfo{author}{\bibfnamefont{D.}~\bibnamefont{Martin-Cano}},
  \bibinfo{author}{\bibfnamefont{L.}~\bibnamefont{Martin-Moreno}},
  \bibinfo{author}{\bibfnamefont{F.~J.} \bibnamefont{Garcia-Vidal}},
  \bibnamefont{and} \bibinfo{author}{\bibfnamefont{E.}~\bibnamefont{Moreno}},
  \bibinfo{journal}{Nano Letters} \textbf{\bibinfo{volume}{10}},
  \bibinfo{pages}{3129} (\bibinfo{year}{2010}).

\bibitem[{\citenamefont{Dzsotjan et~al.}(2011)\citenamefont{Dzsotjan, K\"astel,
  and Fleischhauer}}]{Dzsotjan}
\bibinfo{author}{\bibfnamefont{D.}~\bibnamefont{Dzsotjan}},
  \bibinfo{author}{\bibfnamefont{J.}~\bibnamefont{K\"astel}}, \bibnamefont{and}
  \bibinfo{author}{\bibfnamefont{M.}~\bibnamefont{Fleischhauer}},
  \bibinfo{journal}{Phys. Rev. B} \textbf{\bibinfo{volume}{84}},
  \bibinfo{pages}{075419} (\bibinfo{year}{2011}).

\bibitem[{\citenamefont{Gonzalez-Tudela
  et~al.}(2011)\citenamefont{Gonzalez-Tudela, Martin-Cano, Moreno,
  Martin-Moreno, Tejedor, and Garcia-Vidal}}]{gonzaleztudela11a}
\bibinfo{author}{\bibfnamefont{A.}~\bibnamefont{Gonzalez-Tudela}},
  \bibinfo{author}{\bibfnamefont{D.}~\bibnamefont{Martin-Cano}},
  \bibinfo{author}{\bibfnamefont{E.}~\bibnamefont{Moreno}},
  \bibinfo{author}{\bibfnamefont{L.}~\bibnamefont{Martin-Moreno}},
  \bibinfo{author}{\bibfnamefont{C.}~\bibnamefont{Tejedor}}, \bibnamefont{and}
  \bibinfo{author}{\bibfnamefont{F.~J.} \bibnamefont{Garcia-Vidal}},
  \bibinfo{journal}{Phys. Rev. Lett.} \textbf{\bibinfo{volume}{106}},
  \bibinfo{pages}{020501} (\bibinfo{year}{2011}).

\bibitem[{\citenamefont{Martin-Cano et~al.}(2011)\citenamefont{Martin-Cano,
  Gonzalez-Tudela, Martin-Moreno, Garcia-Vidal, Tejedor, and
  Moreno}}]{martincano11}
\bibinfo{author}{\bibfnamefont{D.}~\bibnamefont{Martin-Cano}},
  \bibinfo{author}{\bibfnamefont{A.}~\bibnamefont{Gonzalez-Tudela}},
  \bibinfo{author}{\bibfnamefont{L.}~\bibnamefont{Martin-Moreno}},
  \bibinfo{author}{\bibfnamefont{F.~J.} \bibnamefont{Garcia-Vidal}},
  \bibinfo{author}{\bibfnamefont{C.}~\bibnamefont{Tejedor}}, \bibnamefont{and}
  \bibinfo{author}{\bibfnamefont{E.}~\bibnamefont{Moreno}},
  \bibinfo{journal}{Phys. Rev. B} \textbf{\bibinfo{volume}{84}},
  \bibinfo{pages}{235306} (\bibinfo{year}{2011}).

\bibitem[{\citenamefont{Bellessa et~al.}(2004)\citenamefont{Bellessa, Bonnand,
  Plenet, and Mugnier}}]{bellessa04a}
\bibinfo{author}{\bibfnamefont{J.}~\bibnamefont{Bellessa}},
  \bibinfo{author}{\bibfnamefont{C.}~\bibnamefont{Bonnand}},
  \bibinfo{author}{\bibfnamefont{J.~C.} \bibnamefont{Plenet}},
  \bibnamefont{and} \bibinfo{author}{\bibfnamefont{J.}~\bibnamefont{Mugnier}},
  \bibinfo{journal}{Phys. Rev. Lett.} \textbf{\bibinfo{volume}{93}},
  \bibinfo{pages}{036404} (\bibinfo{year}{2004}).

\bibitem[{\citenamefont{Dintinger et~al.}(2005)\citenamefont{Dintinger, Klein,
  Bustos, Barnes, and Ebbesen}}]{dintinger05a}
\bibinfo{author}{\bibfnamefont{J.}~\bibnamefont{Dintinger}},
  \bibinfo{author}{\bibfnamefont{S.}~\bibnamefont{Klein}},
  \bibinfo{author}{\bibfnamefont{F.}~\bibnamefont{Bustos}},
  \bibinfo{author}{\bibfnamefont{W.~L.} \bibnamefont{Barnes}},
  \bibnamefont{and} \bibinfo{author}{\bibfnamefont{T.~W.}
  \bibnamefont{Ebbesen}}, \bibinfo{journal}{Phys. Rev. B}
  \textbf{\bibinfo{volume}{71}}, \bibinfo{pages}{035424}
  (\bibinfo{year}{2005}).

\bibitem[{\citenamefont{Salomon et~al.}(2009)\citenamefont{Salomon, Genet, and
  Ebbesen}}]{Salomon09}
\bibinfo{author}{\bibfnamefont{A.}~\bibnamefont{Salomon}},
  \bibinfo{author}{\bibfnamefont{C.}~\bibnamefont{Genet}}, \bibnamefont{and}
  \bibinfo{author}{\bibfnamefont{T.~W.} \bibnamefont{Ebbesen}},
  \bibinfo{journal}{Angew. Chem. Int. Ed.} \textbf{\bibinfo{volume}{48}},
  \bibinfo{pages}{8748} (\bibinfo{year}{2009}).

\bibitem[{\citenamefont{Schwartz et~al.}(2011)\citenamefont{Schwartz,
  Hutchison, Genet, and Ebbesen}}]{Schwartz}
\bibinfo{author}{\bibfnamefont{T.}~\bibnamefont{Schwartz}},
  \bibinfo{author}{\bibfnamefont{J.~A.} \bibnamefont{Hutchison}},
  \bibinfo{author}{\bibfnamefont{C.}~\bibnamefont{Genet}}, \bibnamefont{and}
  \bibinfo{author}{\bibfnamefont{T.~W.} \bibnamefont{Ebbesen}},
  \bibinfo{journal}{Phys. Rev. Lett.} \textbf{\bibinfo{volume}{106}},
  \bibinfo{pages}{196405} (\bibinfo{year}{2011}).

\bibitem[{\citenamefont{Guebrou et~al.}(2012)\citenamefont{Guebrou, Symonds,
  Homeyer, Planet, Garstein, Agramovich, and Bellesa}}]{Aberra12}
\bibinfo{author}{\bibfnamefont{S.~A.} \bibnamefont{Guebrou}},
  \bibinfo{author}{\bibfnamefont{C.}~\bibnamefont{Symonds}},
  \bibinfo{author}{\bibfnamefont{E.}~\bibnamefont{Homeyer}},
  \bibinfo{author}{\bibfnamefont{J.~C.} \bibnamefont{Planet}},
  \bibinfo{author}{\bibfnamefont{Y.~N.} \bibnamefont{Garstein}},
  \bibinfo{author}{\bibfnamefont{V.~M.} \bibnamefont{Agramovich}},
  \bibnamefont{and} \bibinfo{author}{\bibfnamefont{J.}~\bibnamefont{Bellesa}},
  \bibinfo{journal}{Phys. Rev. Lett.} \textbf{\bibinfo{volume}{108}},
  \bibinfo{pages}{066401} (\bibinfo{year}{2012}).

\bibitem[{\citenamefont{Hakala et~al.}(2009)\citenamefont{Hakala, Toppari,
  Kuzyk, Pettersson, Tikkanen, Kunttu, and T\"orm\"a}}]{hakala09a}
\bibinfo{author}{\bibfnamefont{T.~K.} \bibnamefont{Hakala}},
  \bibinfo{author}{\bibfnamefont{J.~J.} \bibnamefont{Toppari}},
  \bibinfo{author}{\bibfnamefont{A.}~\bibnamefont{Kuzyk}},
  \bibinfo{author}{\bibfnamefont{M.}~\bibnamefont{Pettersson}},
  \bibinfo{author}{\bibfnamefont{H.}~\bibnamefont{Tikkanen}},
  \bibinfo{author}{\bibfnamefont{H.}~\bibnamefont{Kunttu}}, \bibnamefont{and}
  \bibinfo{author}{\bibfnamefont{P.}~\bibnamefont{T\"orm\"a}},
  \bibinfo{journal}{Phys. Rev. Lett.} \textbf{\bibinfo{volume}{103}},
  \bibinfo{pages}{053602} (\bibinfo{year}{2009}).

\bibitem[{\citenamefont{Vasa et~al.}(2008)\citenamefont{Vasa, Pomraenke,
  Schwieger, Mazur, Kunets, Srinivasan, Johnson, Kihm, Kim, Runge
  et~al.}}]{vasa08a}
\bibinfo{author}{\bibfnamefont{P.}~\bibnamefont{Vasa}},
  \bibinfo{author}{\bibfnamefont{R.}~\bibnamefont{Pomraenke}},
  \bibinfo{author}{\bibfnamefont{S.}~\bibnamefont{Schwieger}},
  \bibinfo{author}{\bibfnamefont{Y.~I.} \bibnamefont{Mazur}},
  \bibinfo{author}{\bibfnamefont{V.}~\bibnamefont{Kunets}},
  \bibinfo{author}{\bibfnamefont{P.}~\bibnamefont{Srinivasan}},
  \bibinfo{author}{\bibfnamefont{E.}~\bibnamefont{Johnson}},
  \bibinfo{author}{\bibfnamefont{J.~E.} \bibnamefont{Kihm}},
  \bibinfo{author}{\bibfnamefont{D.~S.} \bibnamefont{Kim}},
  \bibinfo{author}{\bibfnamefont{E.}~\bibnamefont{Runge}},
  \bibnamefont{et~al.}, \bibinfo{journal}{Phys. Rev. Lett.}
  \textbf{\bibinfo{volume}{101}}, \bibinfo{pages}{116801}
  (\bibinfo{year}{2008}).

\bibitem[{\citenamefont{Bellessa et~al.}(2008)\citenamefont{Bellessa, Symonds,
  Meynaud, Plenet, Cambril, Miard, Ferlazzo, and Lemaitre}}]{bellesa08}
\bibinfo{author}{\bibfnamefont{J.}~\bibnamefont{Bellessa}},
  \bibinfo{author}{\bibfnamefont{C.}~\bibnamefont{Symonds}},
  \bibinfo{author}{\bibfnamefont{C.}~\bibnamefont{Meynaud}},
  \bibinfo{author}{\bibfnamefont{J.~C.} \bibnamefont{Plenet}},
  \bibinfo{author}{\bibfnamefont{E.}~\bibnamefont{Cambril}},
  \bibinfo{author}{\bibfnamefont{A.}~\bibnamefont{Miard}},
  \bibinfo{author}{\bibfnamefont{L.}~\bibnamefont{Ferlazzo}}, \bibnamefont{and}
  \bibinfo{author}{\bibfnamefont{A.}~\bibnamefont{Lemaitre}},
  \bibinfo{journal}{Phys. Rev. B} \textbf{\bibinfo{volume}{78}},
  \bibinfo{pages}{205326} (\bibinfo{year}{2008}).

\bibitem[{\citenamefont{Gomez et~al.}(2010)\citenamefont{Gomez, Vernon,
  Mulvaney, and Davis}}]{Gomez10}
\bibinfo{author}{\bibfnamefont{D.~E.} \bibnamefont{Gomez}},
  \bibinfo{author}{\bibfnamefont{K.~C.} \bibnamefont{Vernon}},
  \bibinfo{author}{\bibfnamefont{P.}~\bibnamefont{Mulvaney}}, \bibnamefont{and}
  \bibinfo{author}{\bibfnamefont{T.~J.} \bibnamefont{Davis}},
  \bibinfo{journal}{Nano Letters} \textbf{\bibinfo{volume}{10}},
  \bibinfo{pages}{274} (\bibinfo{year}{2010}).

\bibitem[{\citenamefont{Palik}(1985)}]{Palik}
\bibinfo{author}{\bibfnamefont{E.}~\bibnamefont{Palik}},
  \emph{\bibinfo{title}{Handbook of Optical Constants of Solids}}
  (\bibinfo{publisher}{Academic Press Handbook Series (New York), edited by
  Edward D. Palik}, \bibinfo{year}{1985}).

\bibitem[{\citenamefont{Barnes}(1998)}]{barnes98a}
\bibinfo{author}{\bibfnamefont{W.}~\bibnamefont{Barnes}}, \bibinfo{journal}{J.
  Mod. Opt.} \textbf{\bibinfo{volume}{45}}, \bibinfo{pages}{661}
  (\bibinfo{year}{1998}).

\bibitem[{\citenamefont{Tame et~al.}(2008)\citenamefont{Tame, Lee, Lee,
  Ballester, Paternostro, Zayats, and Kim}}]{Zayats}
\bibinfo{author}{\bibfnamefont{M.~S.} \bibnamefont{Tame}},
  \bibinfo{author}{\bibfnamefont{C.}~\bibnamefont{Lee}},
  \bibinfo{author}{\bibfnamefont{J.}~\bibnamefont{Lee}},
  \bibinfo{author}{\bibfnamefont{D.}~\bibnamefont{Ballester}},
  \bibinfo{author}{\bibfnamefont{M.}~\bibnamefont{Paternostro}},
  \bibinfo{author}{\bibfnamefont{A.~V.} \bibnamefont{Zayats}},
  \bibnamefont{and} \bibinfo{author}{\bibfnamefont{M.~S.} \bibnamefont{Kim}},
  \bibinfo{journal}{Phys. Rev. Lett.} \textbf{\bibinfo{volume}{101}},
  \bibinfo{pages}{190504} (\bibinfo{year}{2008}).

\bibitem[{\citenamefont{Archambault et~al.}(2010)\citenamefont{Archambault,
  Marquier, Greffet, and Arnold}}]{Greffet}
\bibinfo{author}{\bibfnamefont{A.}~\bibnamefont{Archambault}},
  \bibinfo{author}{\bibfnamefont{F.}~\bibnamefont{Marquier}},
  \bibinfo{author}{\bibfnamefont{J.-J.} \bibnamefont{Greffet}},
  \bibnamefont{and} \bibinfo{author}{\bibfnamefont{C.}~\bibnamefont{Arnold}},
  \bibinfo{journal}{Phys. Rev. B} \textbf{\bibinfo{volume}{82}},
  \bibinfo{pages}{035411} (\bibinfo{year}{2010}).

\bibitem[{\citenamefont{Fidder et~al.}(1990)\citenamefont{Fidder, Knoester, and
  Wiersma}}]{Fidder90}
\bibinfo{author}{\bibfnamefont{H.}~\bibnamefont{Fidder}},
  \bibinfo{author}{\bibfnamefont{J.}~\bibnamefont{Knoester}}, \bibnamefont{and}
  \bibinfo{author}{\bibfnamefont{D.~A.} \bibnamefont{Wiersma}},
  \bibinfo{journal}{Chemical Physics Letters} \textbf{\bibinfo{volume}{171}},
  \bibinfo{pages}{529 } (\bibinfo{year}{1990}).

\bibitem[{\citenamefont{Carmichael}(1999)}]{carmichael02}
\bibinfo{author}{\bibfnamefont{H.}~\bibnamefont{Carmichael}},
  \emph{\bibinfo{title}{Statistical Methods in Quantum Optics I}}
  (\bibinfo{publisher}{Springer}, \bibinfo{year}{1999}).

\bibitem[{\citenamefont{Laussy et~al.}(2008)\citenamefont{Laussy, del Valle,
  and Tejedor}}]{laussy08a}
\bibinfo{author}{\bibfnamefont{F.~P.} \bibnamefont{Laussy}},
  \bibinfo{author}{\bibfnamefont{E.}~\bibnamefont{del Valle}},
  \bibnamefont{and} \bibinfo{author}{\bibfnamefont{C.}~\bibnamefont{Tejedor}},
  \bibinfo{journal}{Phys. Rev. Lett.} \textbf{\bibinfo{volume}{101}},
  \bibinfo{pages}{083601} (\bibinfo{year}{2008}).

\bibitem[{\citenamefont{Loudon}(2000)}]{loudon_book00a}
\bibinfo{author}{\bibfnamefont{R.}~\bibnamefont{Loudon}},
  \emph{\bibinfo{title}{{The quantum theory of light}}}
  (\bibinfo{publisher}{Oxford Science Publications}, \bibinfo{year}{2000}),
  \bibinfo{edition}{3rd} ed.

\bibitem[{\citenamefont{Ambrose et~al.}(1994)\citenamefont{Ambrose, Goodwin,
  Keller, Martin et~al.}}]{ambrose94a}
\bibinfo{author}{\bibfnamefont{W.~P.} \bibnamefont{Ambrose}},
  \bibinfo{author}{\bibfnamefont{P.~M.} \bibnamefont{Goodwin}},
  \bibinfo{author}{\bibfnamefont{R.~A.} \bibnamefont{Keller}},
  \bibinfo{author}{\bibfnamefont{J.~C.} \bibnamefont{Martin}},
  \bibnamefont{et~al.}, \bibinfo{journal}{Science (New York, NY)}
  \textbf{\bibinfo{volume}{265}}, \bibinfo{pages}{364} (\bibinfo{year}{1994}).

\bibitem[{\citenamefont{Salomon et~al.}(2012)\citenamefont{Salomon, Gordon,
  Prior, Seideman, and Sukharev}}]{Salomon12}
\bibinfo{author}{\bibfnamefont{A.}~\bibnamefont{Salomon}},
  \bibinfo{author}{\bibfnamefont{R.~J.} \bibnamefont{Gordon}},
  \bibinfo{author}{\bibfnamefont{Y.}~\bibnamefont{Prior}},
  \bibinfo{author}{\bibfnamefont{T.}~\bibnamefont{Seideman}}, \bibnamefont{and}
  \bibinfo{author}{\bibfnamefont{M.}~\bibnamefont{Sukharev}},
  \bibinfo{journal}{Phys. Rev. Lett.} \textbf{\bibinfo{volume}{109}},
  \bibinfo{pages}{073002} (\bibinfo{year}{2012}).

\bibitem[{\citenamefont{Mandel and Wolf}(1995)}]{mandel_book95a}
\bibinfo{author}{\bibfnamefont{L.}~\bibnamefont{Mandel}} \bibnamefont{and}
  \bibinfo{author}{\bibfnamefont{E.}~\bibnamefont{Wolf}},
  \emph{\bibinfo{title}{{Optical coherence and quantum optics}}}
  (\bibinfo{publisher}{Cambridge University Press}, \bibinfo{year}{1995}).

\bibitem[{\citenamefont{Ciuti et~al.}(2000)\citenamefont{Ciuti, Schwendimann,
  Deveaud, and Quattropani}}]{ciuti00a}
\bibinfo{author}{\bibfnamefont{C.}~\bibnamefont{Ciuti}},
  \bibinfo{author}{\bibfnamefont{P.}~\bibnamefont{Schwendimann}},
  \bibinfo{author}{\bibfnamefont{B.}~\bibnamefont{Deveaud}}, \bibnamefont{and}
  \bibinfo{author}{\bibfnamefont{A.}~\bibnamefont{Quattropani}},
  \bibinfo{journal}{Phys. Rev. B} \textbf{\bibinfo{volume}{62}},
  \bibinfo{pages}{R4825} (\bibinfo{year}{2000}).

\bibitem[{\citenamefont{Holstein and Primakoff}(1940)}]{holstein40a}
\bibinfo{author}{\bibfnamefont{T.}~\bibnamefont{Holstein}} \bibnamefont{and}
  \bibinfo{author}{\bibfnamefont{H.}~\bibnamefont{Primakoff}},
  \bibinfo{journal}{Phys. Rev.} \textbf{\bibinfo{volume}{58}},
  \bibinfo{pages}{1098} (\bibinfo{year}{1940}).

\bibitem[{\citenamefont{Ciuti et~al.}(2001)\citenamefont{Ciuti, Schwendimann,
  and Quattropani}}]{ciuti01a}
\bibinfo{author}{\bibfnamefont{C.}~\bibnamefont{Ciuti}},
  \bibinfo{author}{\bibfnamefont{P.}~\bibnamefont{Schwendimann}},
  \bibnamefont{and}
  \bibinfo{author}{\bibfnamefont{A.}~\bibnamefont{Quattropani}},
  \bibinfo{journal}{Phys. Rev. B} \textbf{\bibinfo{volume}{63}},
  \bibinfo{pages}{041303} (\bibinfo{year}{2001}).

\bibitem[{\citenamefont{Verger et~al.}(2006)\citenamefont{Verger, Ciuti, and
  Carusotto}}]{verger06a}
\bibinfo{author}{\bibfnamefont{A.}~\bibnamefont{Verger}},
  \bibinfo{author}{\bibfnamefont{C.}~\bibnamefont{Ciuti}}, \bibnamefont{and}
  \bibinfo{author}{\bibfnamefont{I.}~\bibnamefont{Carusotto}},
  \bibinfo{journal}{Phys. Rev. B} \textbf{\bibinfo{volume}{73}},
  \bibinfo{pages}{193306} (\bibinfo{year}{2006}).

\end{thebibliography}



\end{document}